\documentstyle[12pt]{article}
\def\be{\begin{eqnarray}}
\def\ee{\end{eqnarray}}
\def\ba{\begin{array}}
\def\ea{\end{array}}
\def\G{{\cal G}}

\def\a{{\alpha}}
\def\M{{\cal M}}
\def\N{{\cal N}}

\def\S{{\cal S}}
\def\L{{\cal L}}

\def\Z{{\cal Z}}
\def\z{{\it z}}

\def\H{{\cal H}}
\def\C{{\cal C}}
\def\Q{{\cal Q}}
\def\P{{\cal P}}
\def\K{{\cal K}}
\def\Om{\Omega}
\def\om{\omega}
\def\E{{\cal E}}
\def\F{{\cal F}}

\def\P{{\cal P}}
\begin{document}
\begin{center}
{\bf\LARGE {String theory extensions of Einstein-Maxwell fields:\\
         \vskip 3mm
	    the stationary case}}
\end{center}
\vskip 10mm

\begin{center}
{\large Alfredo Herrera-Aguilar\footnote{E-mail address:
{\tt herrera@itzel.ifm.umich.mx}},
Oleg V. Kechkin\footnote{E-mail address: {\tt kechkin@ginette.ifm.umich.mx},\,
{\tt kechkin@depni.npi.msu.su}}
\footnote{On leave from Institute of Nuclear Physics of
M.V. Lomonosov Moscow State University, Vorob'jovy Gory, 119899 Moscow, Russia}}
\end{center}

\vskip 5mm

\begin{center}
Instituto de F\'\i sica y Matem\'aticas\\
Universidad Michoacana de San Nicol\'as de Hidalgo\\
Edificio C--3, Ciudad Universitaria, CP 58040. Morelia, Mich., M\'exico
\end{center}

%\vskip 5mm
%
%\begin{center}
%November 2001
%\end{center}

\vskip 5mm

\begin{abstract}
We present a new approach for generating solutions in
heterotic string theory compactified down to three dimensions
on a torus with $d+n>2$, where $d$ and $n$ stand for the number of
compactified space--time dimensions and Abelian gauge fields,
respectively. It is shown that in the case when $d=2k+1$ and $n$ is
arbitrary, one can apply a solution--generating procedure starting from
solutions of the stationary Einstein theory with $k$
Maxwell fields; our approach leads to classes of solutions which are
invariant with respect to the total group of three-dimensional
charging symmetries.
We consider a particular extension of the stationary
Einstein--multi-Maxwell theory obtained on the basis of
the Kerr--multi-Newman--NUT special class of solutions and establish
the conditions under which the resulting multi--dimensional metric is
free of Dirac string peculiarities.

\end{abstract}
\vskip 5mm
\begin{center}
\end{center}

\newpage

%1111111111111111111111111111111111111111111111111111111111111
\renewcommand{\theequation}{\thesection.\arabic{equation}}
%\newpage
\section{Introduction}
\setcounter{equation}{0}
Symmetry based approaches used for the construction of solutions in the
framework of effective field (low energy) limits of string theories
play an important
role \cite{gsw}, \cite{kir}, \cite{youm}. In this paper we develop a
new approach which allows one to extend the solution space of the
stationary Einstein--multi-Maxwell (EmM) theory to the realm of
heterotic string theory compactified down to three spatial dimensions
on a torus. Namely, we show that a new charging symmetry invariant
subspace of solutions of heterotic string theories (with $d=2k+1$
toroidally compactified space--time dimensions and arbitrary
number $n$ of original Abelian fields) can be generated from the
solution spectrum of the stationary Einstein theory with $k$ Maxwell
fields. In particular, for the critical cases of heterotic ($d=7$) and
bosonic ($d=23$) string theories one must start from the EmM theory
with $k=3$ and $k=11$ Maxwell fields, respectively.

The toroidal compactification of heterotic string theory with arbitrary
values of $d$ and $n$ was originally performed in \cite{hs}--\cite{ms},
whereas the special case when the resulting theory is
three--dimensional was originally studied in
\cite{marcus}--\cite{sen3}. There, the
corresponding symmetric space model was identified and an explicit
representation in terms of a null--curvature matrix was given
(see \cite{bgm} for such models and their classification). In this
paper we exploit the general formalism developed in \cite{mep},
\cite{cs} and \cite{zf} as a natural matrix generalization of the
stationary Einstein--Maxwell theory written down in terms of
potentials which are closely related to the Ernst ones
(see \cite{ernst}, \cite{maz} and \cite{exact}). In the
framework of this formalism, the subgroup of charging symmetry
transformations (which preserve the property of asymptotic flatness of
the solutions) acts as a linear and homogeneous map; this fact allows
one to work with the solution spectrum of the theory in a transparently
charging symmetry invariant form. In particular, all the results of
this paper are automatically invariant with respect to the action of
the total three-dimensional subgroup of charging symmetries of the
heterotic string theory compactified on a torus.

In this paper we continue our investigation on string
theory extensions of Einstein--Maxwell fields. In a previous work
\cite{m1} we have studied two theories with $d+n=2$; here we deal with
theories with $d+n>2$. Such a split of the effective theories with
arbitrary $d$ and $n$, which arises in the low--energy limit of
heterotic string theory, follows from the study of the general
Israel--Wilson--Perj\'es class of solutions of heterotic string theory
performed in \cite{zf}. The new formalism
allows one to construct, in particular, a continuos
generalization of the extremal Israel--Wilson--Perj\'es families of
solutions in the corresponding string theories to the field of
non--extremal ones. In \cite{m1} it was shown that the {\it static}
Einstein--Maxwell theory plays the role of starting system for two
theories with $d+n=2$; in this paper we show that for the theories with
$d+n>2$ such starting systems can be related to the {\it stationary}
Einstein--multi-Maxwell theory. Here we illustrate the developed
general approach by considering an extension of the
Ker--Newman--NUT solution to the realm of heterotic string theory.

%2222222222222222222222222222222222222222222222222222222222222222222
\section{New Formalism for 3D Heterotic String Theory}
\setcounter{equation}{0}

In this section we review the necessary elements of the new formalism
developed in \cite {zf} for the $D$--dimensional ($D=d+3$) heterotic
string theory with $n$ Abelian gauge fields.

We start with the action for the bosonic sector of the low-energy
heterotic string theory \cite{gsw},\cite{kir}:
\be\label{2.1}
\S_D=\int d^DX |{\rm det}G_{MN}|^{\frac{1}{2}}\,e^{-\Phi}
\left ( R_D+\Phi_{,M}\Phi^{,M}-\frac{1}{12}H_{MNK}H^{MNK}-
\frac{1}{4}F^I_{MN}F^{I\,MN}\right ),
\ee
where $H_{MNK}=\partial_MB_{NK}-\frac{1}{2}A^I_MF^I_{NK}+\,{\rm cyclic}\,
\{M,N,K\}$ and $F^I_{MN}=\partial_MA^I_N-\partial_NA^I_M$. Here $X^M$ is the
M-th $(M=1,...,D)$ coordinate of the physical space-time of signature
$(-,+,...,+)$, $G_{MN}$ is the metric, whereas $\Phi$, $B_{MN}$ and $A^I_M$ \,
$(I=1,...,n)$ are the dilaton, Kalb-Ramond and Abelian gauge fields,
respectively. To determine the
result of the toroidal compactification to three dimensions, let us put $D=d+3$,
$X^M=(Y^m,\,x^{\mu})$ with $Y^M=X^m$\, $(m=1,...,d)$ and $x^{\mu}=X^{d+\mu}$
\, $(\mu=1,2,3)$ and introduce the $d\times d$ matrix $G_0={\rm diag}\,
(-1;1,...,1)$,
the $(d+1)\times (d+1)$ and $(d+1+n)\times (d+1+n)$ matrices $\Sigma$ and
$\Xi$ of the form ${\rm diag}\, (-1,-1;1,...,1)$, respectively,
and the $(d+1)\times (d+1+n)$ matrix field $\Z=\Z(x^{\lambda})$
together with the three-metric
$h_{\mu\nu}=h_{\mu\nu}(x^{\lambda})$. In \cite{zf} it was shown that
the resulting theory after the
toroidal compactification of the first $d$ dimensions $Y^m$ can be expressed in terms
of the pair $(\Z,\,h_{\mu\nu})$; its effective dynamics is given by the action
\be\label{2.2}
\S_3=\int d^3x h^{\frac{1}{2}}\left ( -R_3+L_3\right ),
\ee
where $R_3=R_3(h_{\mu\nu)}$ is the curvature scalar for the three-dimensional line
element $ds_3^2=h_{\mu\nu}dx^{\mu}dx^{\nu}$ and
\be\label{2.3}
L_3={\rm Tr}\,\left [\nabla\Z\left (\Xi-\Z^T\Sigma\Z\right )^{-1}
\nabla\Z^T\left (\Sigma-\Z\Xi\Z^T\right )^{-1}\right ].
\ee
In order to translate this ($\sigma$-model) description into the
language of the
field components of the heterotic string theory, let us introduce three doublets of $(\Z,\,h_{\mu\nu})$-related
potentials $(\M_{\alpha},\vec\Omega_{\alpha})$ \, $(\alpha=1,2,3)$ according to the
relations
\be\label{2.4}
&&\M_1=\H^{-1},\quad \nabla\times\vec\Omega_1=\vec J,
\nonumber\\
&&\M_2=\H^{-1}\Z,\quad \nabla\times\vec\Omega_2=\H^{-1}\nabla\Z-\vec J\Z,
\nonumber\\
&&\M_3=\Z^T\H^{-1}\Z,\quad \nabla\times\vec\Omega_3=\nabla\Z^T\H^{-1}\Z-\Z^T\H^{-1}
\nabla\Z+\Z^T\vec J\Z,
\ee
where $\H=\Sigma-\Z\Xi\Z^T$ and $\vec J=\H^{-1}\left (\Z\Xi\nabla\Z^T-\nabla\Z\Xi\Z^T\right )
\H^{-1}$. In Eq. (\ref{2.4}) the scalars $\M_{\alpha}$ are off-shell
defined magnitudes, whereas
the vectors $\vec\Omega_{\alpha}$ are defined on-shell. The scalar and vector potentials
forming any doublet have the same matrix dimensionalities; let us represent them in
the following block form
\be \label{2.5}
\left(
\ba{cc}
1\times 1&1\times d\cr
d\times 1&d\times d
\ea
\right),
\quad
\left(
\ba{ccc}
1\times 1&1\times d&1\times n\cr
d\times 1&d\times d&d\times n
\ea
\right),
\quad
\left(
\ba{ccc}
1\times 1&1\times d&1\times n\cr
d\times 1&d\times d&d\times n\cr
n\times 1&n\times d&n\times n
\ea
\right)
\ee
for $\a=1,2,3$ respectively, where, for example, the
`$13$' block components of the potentials $\M_2$ and $\vec\Omega_2$ are
$1\times n$ matrices. Afterwards let us define the following set
of scalar magnitudes:
\be \label{2.6}
S_{\a}=U_{\a}+S_0^{-1}W_1^TW_{\a},
\ee
where
\be \label{2.6'}
&&S_0=-\M_{1,11}+2\M_{2,11}-\M_{3,11},
\nonumber\\
&&U_1=G_0\M_{1,22}G_0+G_0\M_{2,22}+(\M_{2,22})^TG_0+\M_{3,22},
\nonumber\\
&&U_2=G_0\M_{1,22}-G_0\M_{2,22}G_0+(\M_{2,22})^T-\M_{3,22}G_0,
\nonumber\\
&&U_3=\sqrt 2\left(G_0\M_{2,23}+\M_{3,23}\right),
\nonumber\\
&&W_1=-\M_{1,12}G_0-\M_{2,12}+(\M_{2,21})^TG_0+\M_{3,12},
\nonumber\\
&&W_2=\M_{1,12}-\M_{2,12}G_0+(\M_{2,21})^T-\M_{3,12}G_0,
\nonumber\\
&&W_3=\sqrt2\left (\M_{2,13}+\M_{3,13}\right ),
\ee
and vector fields
\be \label{2.6''}
&&\vec V_1=\left [ -\vec\Omega_{1,12}G_0+\vec\Omega_{2,12}+(\vec\Omega_{2,21})^TG_0
+\vec\Omega_{3,12}\right ]^T,
\nonumber\\
&&\vec V_2=\left [ -\vec\Omega_{1,12}-\vec\Omega_{2,12}G_0+(\vec\Omega_{2,21})^T
-\vec\Omega_{3,12}G_0\right ]^T,
\nonumber\\
&&\vec V_3=\sqrt 2\left ( \vec\Omega_{2,13}+\vec\Omega_{3,13}\right)^T.
\ee

In terms of them the heterotic string theory fields read:
\be\label{2.7}
&&ds^2_D=ds^2_{d+3}=(dY+V_{1\mu}dx^{\mu})^TS_1^{-1}(dY+V_{1\nu}dx^{\nu})+S_0ds_3^2,
\nonumber\\
&&e^{\Phi}=\left | S_0\,{\rm det}\,S_1\right |^{\frac{1}{2}},
\nonumber\\
&&B_{mk}=\frac{1}{2}\left ( S_1^{-1}S_2-S_2^TS_1^{-1}\right )_{mk},
\nonumber\\
&&B_{m\,d+\nu}=\left \{ V_{2\nu}+\frac{1}{2}\left ( S_1^{-1}S_2-
S_2^TS_1^{-1}\right )V_{1\nu}-S_1^{-1}S_3V_{3\nu}\right \}_{m},
\nonumber\\
&&B_{d+\mu\,d+\nu}=\frac{1}{2}\left [ V_{1\mu}^T\left ( S_1^{-1}S_2-
S_2^TS_1^{-1}\right )V_{1\nu}+V_{1\mu}^TV_{2\nu}-V_{1\nu}^TV_{2\mu}\right ],
\nonumber\\
&&A^I_m=\left ( S_1^{-1}S_3\right )_{mI},
\nonumber\\
&&A^I_{d+\mu}=\left ( -V_{3\mu}+S^T_3S^{-1}_1V_{1\mu}\right )_I.
\ee
>From Eq. (\ref{2.7}) it follows that, apart from the magnitudes $S_0$, $S_{\a}$ and
$\vec V_{\a}$, we also must compute ${\rm det}S_1$ and $S_1^{-1}$ in order to obtain
the field components of the heterotic string theory (\ref{2.1}). Therefore, after
some algebraic calculations it can be proved that
\be \label{2.7'}
{\rm det}S_1=\left(1+S_0W_1U_1^{-1}W_1^T\right){\rm det}U_1, \nonumber\\
S_1^{-1}=U_1^{-1}-\frac{S_0^{-1}U_1^{-1}W_1^TW_1U_1^{-1}}{1+S_0^{-1}W_1U_1^{-1}W_1^T}.
\ee
Finally let us point out that the magnitudes $S_0$, $U_{\a}$, $W_{\a}$ and $\vec
V_{\a}$ can be used to explicitly  write down any solution of the theory under
consideration. In Section 4 we shal calculate them for an extension of the stationary
EmM theory to the realm of the low--energy heterotic string theory.

At the end of this Section let us notice that the transformation
\be\label{2.8}
\Z\longrightarrow \C_1\Z\C_2
\ee
with $\C_1^T\Sigma\C_1=\Sigma$ and $\C_2^T\Xi\C_2=\Xi$ is a transparent
symmetry of the theory under consideration (see Eq. (\ref{2.3})). In
\cite{cs} and \cite{zf} it was shown that this symmetry coincides with
the total group of three dimensional charging symmetries. The above
reviewed formalism, based on the use of the matrix potential $\Z$,
possesses the lowest matrix dimensionality ($\Z$ is a
$(d+1)\times(d+n+1)$ matrix field and the theory (\ref{2.3}) is, in
fact, a $O(d+1,d+n+1)/O[(d+1)\times(d+n+1)]$ symmteric space model of
dimension $(d+1)(d+n+1)$ \cite{bgm}). From Eq. (\ref{2.8}) it also
follows that the general transformation of the charging symmetry
subgroup acts as a linear and homogeneous map -- a fact that was
just discussed in the Introduction. It is clear that the new
formalism is specially convenient for the study of asymptotically
flat solutions of heterotic string theory toroidally compactified
to three dimensions because all the results can be obtained in a
transparent charging symmtery invariant form.

%3333333333333333333333333333333333333333333333333333333333333333333333333333
\section{String theories from stationary Einstein--multi-Maxwell system}
\setcounter{equation}{0}

In this section we show how to map solutions of the stationary Einstein
theory with $k$ Maxwell fields into solutions to the three--dimensional
heterotic string theory with $d=2k+1$ toroidally compactified
space--time dimensions and arbitrary number $n$ of Abelian gauge
fields.

First of all let us formulate the main idea of our approach; it is
related to the heterotic string/Einstein--Maxwell theory correspondence
and the explicite form of the Isreal--Wilson--Perj\'es class of
solutions in both of these theories \cite{zf}. In order to achieve this
aim, let us represent the stationary Einstein--Maxwell (EM) theory in a
very similar form to Eq. (\ref{2.3}). Namely, it is well known that the
effective three--dimensional Lagrangian of the stationary EM theory
reads:
\be\label{3.1}
L_3=L_{EM}=\frac{1}{2f^2}\left | \nabla E-\bar F\nabla F\right |^2-
\frac{1}{f}\left | \nabla F\right |^2,
\ee
where $f=\frac{1}{2}(E+\bar E-|F|^2)$, and $E$ and $F$ are the
conventional Ernst potentials. Let us introduce the $1\times 2$ matrix
potential
\be\label{3.2}
z=(z_1\,\,z_2)
\ee
with
\be\label{3.3}
z_1=\frac{1-E}{1+E},\qquad z_2=\frac{\sqrt 2F}{1+E}.
\ee
Then
\be\label{3.4}
L_{EM}=2\frac{\nabla z(\sigma_3-z\dagger z)^{-1}\nabla z\dagger}
{1-z\sigma_3z\dagger},
\ee
where $\sigma_3={\rm diag}\,(1\,-1)$. By comparing Eqs. (\ref{2.3}) and
(\ref{3.4}) it follows that the map
\be\label{3.5}
\Z\longleftrightarrow z,\quad \Xi\longleftrightarrow\sigma_3,\quad
\Sigma\longleftrightarrow 1,
\ee
together with the interchange of operations
$T\longleftrightarrow \dagger$, relates three--dimensional heterotic
string
and stationary Einstein-Maxwell theories; the factor `2' in
(\ref{3.4}) can be understood as a consequence of the exact matrix
representation of complex magnitudes (see below and \cite{zf} as well).
Therefore, the Israel--Wilson-Perj\'es (IWP) class of solutions of
the Einstein-Maxwell theory \cite{iwp} can be rewritten in terms of the
$z$--potential as $z=\lambda q$, where $\lambda=\lambda(x^{\mu})$ is
a complex harmonic function ($\nabla^2\lambda=0$), $q$ is a
$1\times 2$--matrix constant parameter and the parameter
$\kappa=q\sigma_3q\dagger$ vanishes; in this case, the corresponding
three--dimensional metric $h_{\mu\nu}$ is flat. It is clear that, in
view of the correspondence (\ref{3.5}), the IWP class of solutions
of the heterotic string theory arises in the framework of the ansatz
\be\label{3.6}
\Z=\Lambda\Q,
\ee
where $\Lambda=\Lambda(x^{\mu})$ is a real harmonic matrix function and
$\Q$ is a constant matrix parameter. In \cite{zf} it was shown that
this fact actually takes place if the parameter
\be\label{3.7}
\kappa=\Q\Xi\Q^T
\ee
vanishes and the three--metric $h_{\mu\nu}$ is flat again, in complete
accordance with the correspondence (\ref{3.5}). There, it also was
shown that the restriction $\kappa=0$ completely fixes the
dimensionality of the matrices $\Lambda$ and $\Q$: for two theories
with $d+n=2$, such matrices have dimensions $(d+1)\times 1$ and
$1\times 3$, respectively, whereas for the theories with $d+n>2$
the dimensions are $(d+1)\times 2$ and $2\times (d+n+1)$, respectively.

The main idea of our approach is to remove the $\kappa=0$
restriction and, therefore, to consider the resulting generalization
of the IWP class of solutions of heterotic string theory to the
subspace of the non--extremal solutions. This means that we shall
preserve the form of the ansatz (\ref{3.6}) and the dimensions of the
matrices $\Lambda$ and $\Q$, but we shall allow arbitrary values of the
parameter $\kappa$ defined by Eq. (\ref{3.7}). Such a procedure can be
applied in a very natural way in the framework of the stationary EM
theory, where it defines, for instance, a continuous extension of the
extremal Kerr--multi-Newman--NUT solution to the corresponding
non--extremal one \cite{exact}. This extension is really interesting
from the point of view of physical applications; the example given
above concerns black hole physics in EM theory \cite{bh}. In view of
the correspondence (\ref{3.5}) the same motivation for the study of the
ansatz (\ref{3.6}) with $\kappa\ne 0$ must be valid for the low--energy
heterotic string theory; thus, the study of such an ansatz is also
interesting in the framework of black holes in string theory
\cite{youm}, \cite{ap}.

In this paper we consider string theories with $d+n>2$, when $\kappa$
is a symmetric $2\times 2$--matrix. By straightforwardly substituting
the ansatz (\ref{3.6}) into the equations of motion derived from Eqs.
(\ref{2.2}) and (\ref{2.3}), one obtains
\be
\nabla^2\Lambda+2\nabla\Lambda\kappa\Lambda^T\left(\Sigma-
\Lambda\kappa\Lambda^T\right)^{-1}\nabla\Lambda=0,
\nonumber \\
R_{3\,\,\mu\nu}=Tr\left[\Lambda_{,(\mu}\kappa
\left(1-\Lambda^T\Sigma\Lambda\kappa\right)^{-1}\Lambda^T_{,\nu)}
\left(\Sigma-\Lambda\kappa\Lambda^T\right)^{-1}\right].
\label{3.8}
\ee
It is obvious that in the case $\kappa=0$ we recover the extremal case
studied in \cite{zf}, whereas for $\kappa\ne 0$ we have the above
announced continuos extension of the formalism to the non--extremal
case. Below we study the situation when $\kappa$ is nonzero and,
moreover, nondegenerate matrix with signature $\tilde\Sigma={\rm
diag}(-1,-1)$. The reason for considering such a particularization of
the ansatz (\ref{3.6}) is that, in this case, the effective system
defined by Eqs. (\ref{3.8}) corresponds to some new heterotic string
theory by itself. Actually, The Eqs. (\ref{3.8}) are the equations of
motion for the action (\ref{2.2}) with the matter Lagrangian
(\ref{2.3}) replaced by
\be\label{3.9}
\tilde\L_3=Tr\left[\nabla\Lambda\kappa
\left(1-\Lambda^T\Sigma\Lambda\kappa\right)^{-1}\nabla\Lambda^T
\left(\Sigma-\Lambda\kappa\Lambda^T\right)^{-1}\right].
\ee
Then, as an algebraic fact it follows that there exists a nondegenerate
matrix $K$ such that
\be\label{3.10}
\kappa=K\tilde\Sigma K^T.
\ee
Let us introduce the new matrix potential
\be\label{3.11}
\tilde\Z=K^T\Lambda^T
\ee
and set $\tilde\Xi=\Sigma$. We claim that it is possible to rewrite the
effective Lagrangian $\tilde\L_3$ (\ref{3.9}) in terms of  $\tilde\Z$,
$\tilde\Sigma$ and $\tilde\Xi$; the resulting Lagrangian exactly
coincides with the relation (\ref{2.3}) up to the tilde. Thus, the
effective system (\ref{3.9}) is nothing else than the heterotic string
theory with $\tilde d=1$ compactified dimensions and $\tilde n=d-1$
Abelian gauge fields. From Eqs. (\ref{3.6}) and (\ref{3.11}) it follows
that
\be\label{3.12}
\Z=\tilde\Z^TT.
\ee
where $T=K^{-1}\Q$. Eq. (\ref{3.12}) maps the space of solutions of
the theory in terms of
$\tilde\Z$ into that of the theory in terms of the potential $\Z$, so
that the matrix $T$ plays the role of a symmetry
operator. Let
us now calculate the general explicit form of such symmetry operator
using Eqs. (\ref{3.6}) and (\ref{3.10}). Without loss of generality
(see \cite{zf} for details), the
matrix $\Q$ can be parametrized in the form
\be\label{3.13}
\Q=
\left(
\ba{ccc}
1&0&n_1^T\cr
0&1&n_2^T
\ea
\right),
\ee
where $n_a$ ($a=1,2$) are two $(d+n-1)\times 1$ columns. Thus, the
extremal case corresponds to the restriction $n_a^Tn_b=\delta_{ab}$,
i.e. it is realized by the unit orthogonal columns $n_a$. Our
generalization of the extremal ansatz corresponds, in this geometric
language, to the case of columns with arbitrary length and arbitrary
angle between them which is compatible with the signature
$\tilde\Sigma$ of the matrix $\kappa$.

Now we are able to compute the matrix $\kappa$ and to determine the
quantity $K$ using, for example, the orthogonalization procedure of the
theory of quadratic forms. A special solutions reads:
\be\label{3.14}
K=
\left(
\ba{ccc}
\sqrt{1-n_1^Tn_1}&\quad&0\cr
\quad &\quad &\quad \cr
-\frac{n_1^Tn_2}{\sqrt{1-n_1^Tn_1}}&\quad&
\sqrt{\frac{1-n_1^Tn_1-n_2^Tn_2+(n_1^Tn_1)(n_2^Tn_2)-(n_1^Tn_2)^2}
{1-n_1^Tn_1}}
\ea
\right),
\ee
In order to obtain a general solution $K$ to the quadratic equation
(\ref{3.10}) one must generalize the special solution (\ref{3.14})
through the map $K\longrightarrow K\C$ where
$\C^T\tilde\Sigma\C=\tilde\Sigma$, i.e. $\C\in O(2)$. However, this
map is effectively equivalent to the transformation
$\tilde\Z\longrightarrow \C^{T\,\,-1}\tilde\Z$ as it follows from Eq.
(\ref{3.12}). It is clear that $\C^{T\,\,-1}$ is nothing more than an
alternative notation for the `left' subgroup $\C_1$ of the charging
symmetry transformation (see Eq. (\ref{2.8})), thus, it can be omitted
for the charging symmetry invariant classes of solutions repersented by
$\tilde\Z$. Thus, without loss of generality one can take $\Q$ and $K$
in the form given by Eqs. (\ref{3.13}) and (\ref{3.14}) when
constructing a symmetry operator according to Eq. (\ref{3.12}). It
is worth noticing that the definition of $K$ is consistent with the
assumed signature of $\kappa$.

Now let us consider a special situation with $d=2k+1$, when the
potential $\tilde\Z$ can be splitted into $k+1$ $2\times 2$--matrix
blocks:
\be\label{3.15}
\tilde\Z=
\left(
\ba{cccc}
\tilde\Z_1,&
\tilde\Z_2,& ...\,,&
\tilde\Z_{k+1}
\ea
\right).
\ee
Let us consider a consistent ansatz with
\be\label{3.16}
\tilde\Z_{\P}=
\left(
\ba{cc}
z_{\P}'&-z_{\P}'' \cr
z_{\P}''&z_{\P}'
\ea
\right),
\ee
where $\P=1,2,...,k+1$. Let us introduce $k+1$ complex functions
$z_{\P}=z_{\P}'+iz_{\P}''$. Our statement is that the heterotic
string theory field equations corresponding to the special subsystem
(\ref{3.15})--(\ref{3.16}) can be derived from the effective Lagrangian
\be\label{3.17}
L_3=L_{EmM}=2\frac{\nabla\tilde z(\tilde\sigma_3-\tilde z\dagger\tilde
z)^{-1}
\nabla\tilde z\dagger}{1-\tilde z\tilde\sigma_3\tilde z\dagger},
\ee
where
$\tilde z=\left(\tilde z_1,\tilde z_2,\,...\,\tilde z_{k+1}\right)$ and
$\tilde\sigma_3={\rm diag}(1,-1,-1,...,-1)$. It is clear that in
the case $k=1$ one recovers the Eq. (\ref{3.4}), i.e. one deals with
the stationary Einstein--Maxwell theory. It is natural to suppose that
in the case of arbitrary $k$ one obtains the stationary
Einstein theory with $k$ Maxwell fields. In order to verify this
statement it is convenient to introduce new variables (compare to Eq.
(\ref{3.3}))
\be\label{3.18}
E=\frac{1-\tilde z_1}{1+\tilde z_1},\qquad F_p=\frac{\sqrt
2\tilde z_{p+1}}{1+\tilde z_1},
\ee
where $p=1,2,...,k$. In terms of these variables the effective
Lagrangian adopts the form
\be\label{3.19}
L_{EmM}=\frac{1}{2f^2}\left | \nabla E-\bar F_p\nabla F_p\right |^2-
\frac{1}{f}\left | \nabla F_p\right |^2,
\ee
where $f=\frac{1}{2}(E+\bar E-F_p\bar F_p)$, thus, they can be
interpreted as the conventional Ernst potentials of the classical
four--dimensional Einstein theory with $k$ Maxwell fields in the
stationary case.

The theory (\ref{3.17}) can be studied in a form which is very close
to that of the heterotic string theory (see Section 2) by using the
correspondence (\ref{3.5}) modified to the case of $k$ Maxwell fields,
i.e., by exchanging
\be\label{3.20}
z\longrightarrow \tilde z, \qquad
\sigma_3\longrightarrow \tilde\sigma_3,
\ee
(see Eqs. (\ref{3.4}) and
(\ref{3.7})). Namely, it is convenient to introduce the doublets
$(\tilde m_{\a}, \tilde {\vec \omega_{\a}})$ with
\be\label{3.21}
&&\tilde m_1=\tilde h^{-1},\qquad
\nabla\times\tilde{\vec\omega_1}=\tilde{\vec j},
\nonumber \\
&&\tilde m_2=\tilde h^{-1}\tilde z,\qquad
\nabla\times\tilde{\vec\omega_2}=\tilde h^{-1}\nabla\tilde z-
\tilde{\vec j}\tilde z,
\nonumber \\
&&\tilde m_3=\tilde h^{-1}\tilde z\dagger\tilde z,\qquad
\nabla\times\tilde{\vec\omega_3}=\tilde
h^{-1}(\nabla\tilde z\dagger\tilde z-
\tilde z\dagger\nabla\tilde z)+\tilde{\vec j}\tilde z\dagger\tilde z,
\ee
where $\tilde h=-(1-\tilde z\tilde\sigma_3\tilde z\dagger)$ and
$\tilde{\vec j}=-\tilde
h^{-2}\left(\tilde z\tilde\sigma_3\nabla\tilde z\dagger-
\nabla\tilde z\tilde\sigma_3\tilde z\dagger\right)$, (compare with Eq.
(\ref{2.4})). It is clear that the doublet
($\tilde m_1,\tilde{\vec\omega_1}$) consists of complex functions,
whereas the doublets ($\tilde m_2,\tilde{\vec\omega_2}$) and
($\tilde m_3,\tilde{\vec\omega_3}$), of $1\times (k+1)$-- and
$(k+1)\times (k+1)$--matrices, respectively. From Eqs.
(\ref{3.15})--(\ref{3.16})
we extract a rule for reconstructing the heterotic string theory
described by the potential $\tilde\Z$ of dimension $2\times[2(k+1)]$.
Further, from Eqs. (\ref{3.12})--(\ref{3.14}) one obtains the explicit
form of the symmetry map $\tilde\Z\longrightarrow \Z$. It is
interesting that this map is nonholomorphic due to the transposition of
$\tilde\Z$ in Eq.  (\ref{3.12}), which is equivalent to the Hermitean
conjugation of $\tilde z$ in view of the correspondence (\ref{3.20})
discussed above.

Thus, a symmetry transformation that maps the space of solutions
of the stationary Einstein theory with $k$ Maxwell fields into the
corresponding subspace of solutions of the heterotic string with
$d=2k+1$ toroidally compactified dimensions and $n$ arbitrary Abelian
gauge fields is established by the following procedure. First of all,
one must calculate in explicit form three doublets of potentials
$(\tilde m_{\a}, \tilde {\vec \omega_{\a}})$ for the stationary EmM
theory. After that one must rewrite them in the form
$(\tilde M_{\a}, \tilde {\vec\Omega_{\a}})$ using the exact matrix
representation of complex magnitudes (Eqs. (\ref{3.15})--(\ref{3.16})
give, in fact, an example of such a representation of the complex
potential $\tilde z$ in terms of the real potential $\tilde\Z$). The
next step consists of calculating the matrix potentials
$(M_{\a}, \vec\Omega_{\a})$ for the heterotic string theory which
is an image of the EmM system according to the map (\ref{3.12}).
Finally, one must obtain the explicit form of the magnitudes
$S_0$, $W_{\a}$, $U_{\a}$ and $\vec V_{\a}$,
using the found potentials $(M_{\a}, \vec\Omega_{\a})$.

At the end of this Section let us compute the doublets
$(M_{\a}, \vec\Omega_{\a})$. By using Eqs. (\ref{2.4}) and
(\ref{3.12}), after some algebraic calculations, one obtains
\be\label{3.22}
&&M_1=\Sigma+\Sigma\tilde M_3\Sigma,\qquad
\vec\Omega_1=-\Sigma\tilde{\vec\Omega_3}\Sigma,
\nonumber \\
&&M_2=-\Sigma\tilde{M_2^T}T,\qquad
\vec\Omega_2=-\Sigma\tilde{\vec\Omega_2^T}T,
\nonumber \\
&&M_3=T^T(\tilde{M_1}+1)T, \qquad
\vec\Omega_3=-T^T\tilde{\vec\Omega_1}T.
\ee
In the next Section we shall exploit these formulae in order to
construct a subspace of solutions for the heterotic string theory
with $d=2k+1$ and arbitrary $n$ starting from the
stationary Einstein theory with $k$ Maxwell fields.

%%%%%%%%%%%%%%%%%%%%%%%%%%%%%%%%%%%%%%%%%%%%%%%%%%%%%%%%%%%%%%%%%%%%%%%%%%%
%4444444444444444444444444444444444444444444444444444444444444444444444444444
\section{Solutions via Kerr--multi-Newman--NUT family}
\setcounter{equation}{0}

In order to calculate the potentials $S_0$, $W_{\a}$, $U_{\a}$ and
$\vec V_{\a}$ for the heterotic string fields, which correspond to the
Einstein--multi-Maxwell ones, according to the scheme developed in the previous
Section, let us parametrize the potential $\tilde\Z$ and the symmetry operator
$T$ in the appropriate form. For $\tilde\Z$ is convenient to set
\be\label{4.1}
\tilde\Z=\left(\tilde\Z_I\,\tilde\Z_{II}\right),
\ee
where $\tilde\Z_I$ is a $2\times 1$--column and $\tilde\Z_{II}$
is a $2\times(2k+1)$--matrix that read
\be\label{4.2}
\tilde\Z_I=
\left(
\ba{c}
\tilde\z_1^{'} \cr
\tilde\z_1^{''}
\ea
\right),
\qquad
\tilde\Z_{II}=
\left(
\ba{ccc}
-\tilde\z_1^{''} & \tilde\z_{1+p}^{'} & -\tilde\z_{1+p}^{''} \cr
\tilde\z_1^{'} & \tilde\z_{1+p}^{''} & \tilde\z_{1+p}^{'}
\ea
\right).
\ee
Therefore, for $T$ we choose the following segmentation
\be\label{4.3}
T=\left(T_I\,\,T_{II}\,\,T_{III}\right)
\ee
where $T_I$ is a $2\times 1$--column, $T_{II}$ is a
$2\times(2k+1)$--matrix and $T_{III}$ is a
$2\times n$--matrix, i.e.
\be\label{4.4}
&&T_I=\K^{-1}\Q_I, \qquad
\tilde\Q_I=
\left(
\ba{c}
1\cr
0
\ea
\right),
\nonumber \\
&&T_{II}=\K^{-1}\Q_{II}, \qquad
Q_{II}=
\left(
\ba{cc}
0 & r_1^T \cr
1 & r_2^T
\ea
\right),
\nonumber \\
&&T_{III}=\K^{-1}\Q_{III}, \qquad
\tilde\Q_{III}=
\left(
\ba{c}
l_1^T\cr
l_2^T
\ea
\right),
\ee
where we have naturally decomposed
$n_a^T=\left(r_a^T\,\,l_a^T\right)$,
into the rows $r_a^T$ and $l_a^T$ of dimension
$1\times 2k$ and $1\times n$, respectively. It is worth noticing that
the block representation (\ref{4.1}) and (\ref{4.3}) of the magnitudes
$\tilde\Z$ and $T$ is directly related to the number of compactified
dimensions and Abelian vector fields ($2k+1$ and $n$, respectively)
and, thus, is actually fruitful for the application of Eqs (\ref{2.7}).
Thus, after some algebraic computations, Eqs. (\ref{2.7}) yield the
following expression for the scalar $S_0$
\be\label{4.5}
S_0=1-T_I^TT_I-\tilde h^{-1}\left(\tilde\Z_I-T_I\right)^T
\left(\tilde\Z_I-T_I\right),
\ee
whereas for the columns $W_{\a}$ one obtains:
\be\label{4.6}
&&W_1=T_I^TT_{II}+\tilde h^{-1}
\left(\tilde\Z_I-T_I\right)^T\left(\tilde\Z_{II}-T_{II}\right),
\nonumber \\
&&W_2=-\left[T_I^TT_{II}+\tilde h^{-1}
\left(\tilde\Z_I+T_I\right)^T\left(\tilde\Z_{II}+T_{II}\right)\right]G_0,
\nonumber \\
&&W_3=\sqrt2\left[T_I+\tilde h^{-1}
\left(\tilde\Z_I+T_I\right)\right]^TT_{III};
\ee
finally, for the matrix potentials $U_{\a}$ one gets
\be\label{4.7}
&&U_1=G_0+T_{II}^TT_{II}+\tilde h^{-1}
\left(\tilde\Z_{II}-T_{II}\right)^T\left(\tilde\Z_{II}-T_{II}\right),
\nonumber \\
&&U_2=-\left[T_{II}^TT_{II}-\tilde h^{-1}
\left(\tilde\Z_{II}-T_{II}\right)^T\left(\tilde\Z_{II}+T_{II}\right)\right]G_0,
\nonumber \\
&&U_3=\sqrt2\left[T_{II}-\tilde h^{-1}
\left(\tilde\Z_{II}-T_{II}\right)\right]^TT_{III}.
\ee
On the other hand, the vector matrices $\tilde{\vec\Om}_{\a}$
possess the following parametrization
\be\label{4.8}
\tilde{\vec\Om_1}=\vec\om\epsilon,
\quad\quad
\tilde{\vec\Om_2}=\left(\tilde{\vec\Om}_{2,I}\,\,\tilde{\vec\Om}_{2,II}\right),
\quad\quad
\tilde{\vec\Om_3}=
\left(
\ba{cc}
0 & \tilde{\vec\Om}_{3,I\,\,II} \cr
\quad & \quad \cr
\tilde{\vec\Om}_{3,II\,\,I} &
\tilde{\vec\Om}_{3,II\,\,II}
\ea
\right),
\ee
where $\vec\om=\tilde{\vec\om}_1^{''}$ (the magnitude $\tilde{\vec\om}_1$
is purely imaginary, i.e. $\tilde{\vec\om}_1^{'}=0$), $\epsilon$ is the
antisymmetric $2\times2$--matrix with $\epsilon_{12}=-1$, the block
componenets of $\tilde{\vec\Om}_2$ read
\be\label{4.9}
\tilde{\vec\Om}_{2,I}=
\left(
\ba{c}
\tilde{\vec\om}_{2,1}^{'}  \cr
\quad \cr
\tilde{\vec\om}_{2,1}^{''}
\ea
\right), \qquad\qquad
\tilde{\vec\Om}_{2,II}=
\left(
\ba{ccc}
-\tilde{\vec\om}_{2,1}^{''} & \tilde{\vec\om}_{2,1+p}^{'} &
-\tilde{\vec\om}_{2,1+p}^{''} \cr
\quad &\quad &\quad \cr
\tilde{\vec\om}_{2,1}^{'} & \tilde{\vec\om}_{2,1+p}^{''} &
\tilde{\vec\om}_{2,1+p}^{'}
\ea
\right),
\ee
whereas
$\tilde{\vec\Om}_{3,II\,\,I}=-\tilde{\vec\Om}^T_{3,I\,\,II}$,
$\tilde{\vec\Om}_{3,II\,\,II}=-\tilde{\vec\Om}^T_{3,II\,\,II}$
$(\tilde{\vec\Om}_3=-\tilde{\vec\Om}^T_3)$ and
\be\label{4.10}
\tilde{\vec\Om}_{3,I\,\,II}=
\left(
\ba{cccccc}
-\tilde{\vec\om}_{3,1\,\,1}^{''} & \tilde{\vec\om}_{3,1\,\,1+p}^{'} &
-\tilde{\vec\om}_{3,1\,\,1+p}^{''}
\ea
\right).
\ee
Note that, in view of Eqs. (\ref{2.8}) and (\ref{3.22}), only the
magnitudes  $\tilde{\vec\om}$, $\tilde{\vec\Om}_{2,I}$,
$\tilde{\vec\Om}_{2,II}$ and $\tilde{\vec\Om}_{3,I\,\,II}$ are
necessary for the explicit construction of the potential $\vec V_{\a}$.
Finally, the explicite expressions for these vectors read
\be\label{4.11}
&&\vec V_1=\tilde{\vec\om}T_{II}^T\epsilon T_{I}-\tilde{\vec\Om}^T_{2,II}T_I+
T_{II}^T\tilde{\vec\Om}_{2,I}-\tilde{\vec\Om}^T_{3,I\,\,II},
\nonumber \\
&&\vec V_2=-\G_0\left(\tilde{\vec\om}T_{II}^T\epsilon T_{I}+\tilde{\vec\Om}^T_{2,II}T_I+
T_{II}^T\tilde{\vec\Om}_{2,I}+\tilde{\vec\Om}^T_{3,I\,\,II}\right),
\nonumber \\
&&\vec V_3=\sqrt2\left(\tilde{\vec\om}\epsilon T_{I}+\tilde{\vec\Om}_{2,I}\right)^TT_{III}.
\ee
At this stage some remarks are in order: Eqs. (\ref{4.5}), (\ref{4.7})
and (\ref{4.11}) also possess another parametrization which is based on
the identities $\K\K^T=-\kappa$ and
${\K}\epsilon\K^T=\sqrt{\rm det \kappa}\epsilon$. In both
representations the potentials $S_0$, $W_{\a}$, $U_{\a}$ and $\vec V_{\a}$
become trivial ($S_0=1$, $U_1=G_0$, other fields vanish) for a starting
Einstein--multi-Maxwell solution corresponding to $\tilde\Z=0$.
This fact reflects the underlying property of the primordial symmetry
map (\ref{3.12}) in the language of the potentials which define the
components of the physical fields of string theory. As a last remark
let us point out that the appearance of the $(2k+1)\times(2k+1)$--matrix
$G_0={\rm diag}(-1,1,1,...,1)$, which describes the flat metric
corresponding to extra dimensions, in Eq. (\ref{4.7}) is very natural.

As a matter of fact, our generating--procedure, based on the use
of Eq. (\ref{3.12}) and the special choice of the starting matrix
potential $\tilde\Z$ in an Einstein--multi-Maxwell form (see Eqs.
(\ref{4.1}) and (\ref{4.2})), brakes the complex structure of the
starting theory. Actually, in the general case, the
$2\times[2(k+1)+n]$--dimensional symmetry operator $T$ does not
represent any complex magnitude $t$ of dimension $1\times(k+1+n/2)$.
In particular, the number of Abelian gauge fields $n$ can be even.
However, if $n$ is odd, i.e., $n=2J$, and also
$r_{2,2p}=r_{1,2p-1}\equiv r^{'}_p$,
$r_{2,2p-1}=-r_{1,2p}\equiv r^{''}_p$,
$l_{2,2j}=l_{1,2j-1}\equiv l^{'}_j,$
$l_{2,2j-1}=-l_{1,2j}\equiv l^{''}_j$, $(p=1,2,...,k; j=1,2,...,J)$,
then
\be\label{4.12}
T=\frac{1}{\sqrt{1-\N^2}}
\left(
\ba{cccccc}
1 & 0 &r^{'}_p & -r^{''}_p & l^{'}_j & -l^{''}_j \cr
\quad & \quad & \quad & \quad & \quad & \quad  \cr
0 & 1 &r^{''}_p & r^{'}_p & l^{''}_j & l^{'}_j
\ea
\right),
\ee
where $\N^2=n_1^Tn_1=n_2^Tn_2$. In this special case, vectors $n_1$
and $n_2$ have the same length and are mutually orthogonal
($n_1^Tn_2=0$). Therefore, from Eq. (\ref{4.12}) it immediately follows
that the operator $T$ is a real matrix representation of the complex
$1\times(k+1+J)$ row $t=|1-\N^2|^{-1/2}\left(1\,\,r_p\,\,l_j\right)$
where $r_p=r^{'}_p+ir^{''}_p$ and $l_j=l^{'}_j+il^{''}_j$. Notice that
in the special case under consideration it is possible to express Eq.
(\ref{3.22}) in a complex form by substituting
$M_{\a}\longrightarrow m_{\a}$, $\tilde{M}_{\a}\longrightarrow
\tilde{m}_{\a}$, $T\longrightarrow t$ and
$\Sigma\longrightarrow\tilde\sigma_3$. Thus, in this special case it is
possible to keep pure complex notations.

As an example of a concrete class of solutions of the
Einstein--multi-Maxwell theory one can consider the solution which
arises in the framework of the ansatz
\be\label{4.13}
\tilde\z=\lambda \tilde q,
\ee
where $\lambda$ is a complex function and $\tilde q$ is a $1\times
(k+1)$ constant complex row. The corresponding effective system is
related to the Lagrangian
\be\label{4.14}
L_3=2\tilde\kappa\frac{|\nabla\lambda|^2}
{\left(1-\tilde\kappa|\lambda|^2\right)},
\ee
in the case of $\tilde\kappa \ne 0$, where
\be\label{4.15}
\tilde\kappa=\tilde q\tilde\sigma_3\tilde q\dagger.
\ee
When $\tilde\kappa =0$ one obtains a decoupled three--dimensional flat
space and a harmonic field $\lambda$. The parameter $\tilde\kappa$
plays the role of a coupling constant between three--dimensional
gravity and the complex scalar field $\lambda$. We claim that the
following concrete choice of the scalar field and the three metric
\be\label{4.16}
\lambda=\frac{1}{R-ia\cos\theta},
\nonumber
\ee
\be
ds^2_3=\Delta\left(\frac{dR^2}{R^2+a^2-\tilde\kappa}+d\theta^2\right)+
\left(R^2+a^2-\tilde\kappa\right)\sin^2\theta d\varphi^2,
\ee
where $\Delta=R^2+a^2\cos^2\theta-\tilde\kappa$ and $a$ is a constant,
gives a solution of the corresponding equations of motion. Note that in
Eq. (\ref{4.16}) the value of the parameter $\tilde\kappa$ is
arbitrary. In what follows, this concrete class of solutions will be
considered as the typical starting one in the framework of the
developed solution--generating scheme.

To start with, we need explicit expressions for the magnitudes
$\tilde m_{\a}$ and $\tilde{\vec\om}_{\a}$. For the scalar sector one
immediately gets
\be\label{4.17}
&&\tilde m_1=-\left(1+\frac{\tilde\kappa}{\Delta}\right),
\nonumber \\
&&\tilde m_2=-\tilde q\frac{R+ia\cos\theta}{\Delta},
\nonumber \\
&&\tilde m_3=-\tilde q\dagger\tilde q\frac{1}{\Delta},
\ee
whereas for the vector one, after the corresponding integration, one
finds that
\be\label{4.18}
&&\tilde{\vec\om}_{1\,\varphi}=-i\frac{a\tilde\kappa\sin^2\theta}{\Delta},
\nonumber \\
&&\tilde{\vec\om}_{2\,\varphi}=\tilde
q\left(-\cos\theta+ia\sin^2\theta\frac{R+ia\cos\theta}{\Delta}\right),
\nonumber \\
&&\tilde{\vec\om}_{3\,\varphi}=-i\tilde q\dagger\tilde
q\frac{a\sin^2\theta}{\Delta},
\ee
and other vector components vanish.
These relations defines the extension of the solution (\ref{4.16}) to
the realm of the heterotic string theory according to the relations
(\ref{2.6}), (\ref{2.9}), (\ref{4.5})--(\ref{4.7}) and (\ref{4.11}).
Let us discuss
on both, the starting Einstein--multi-Maxwell family of solutions and
the resulting heterotic string theory fields.

First of all, let us compute the Ernst potentials (\ref{3.18})
corresponding to the solution (\ref{4.16}):
\be\label{4.20}
&&\E=1-\frac{2(M+iN)}{r+i(N-a\cos\theta)},
\nonumber \\
&&\F_p=\sqrt2\frac{\left(e_p+ig_p\right)}
{r+i(N-a\cos\theta)},
\ee
where
\be\label{4.19}
\tilde q_1=M+iN, \qquad \quad
\tilde q_{1+p}=e_p+ig_p
\ee
and $r=R+M$ and $\tilde\kappa=-M^2-N^2+\sum_p\left(e^2_p+g^2_p\right)$.
It is clear that our starting solution is precisely
the Kerr--multi-Newman family of solutions with non--trivial NUT
parameter. Thus, $(r,\theta, \varphi)$ stand for conventional
oblate spheroidal coordinates, whereas the parameters $M$, $N$,
$e_p$ and $g_p$ are the mass, NUT, electric and magnetic charges,
respectively. Another interesting issue concerns the asymptotical
flatness of the resulting multidimensional field configuration in the
framework of our solution--generating method in the general case. It
turns out that the generating field configurations contain the
so--called ``Dirac strings" and are not asymptotically flat, i.e. a
term which is proportional to $\cos\theta$ at spatial infinity
($R\longrightarrow\infty$). The same situation takes place for the
starting four--dimensional metric of the Einstein--multi-Maxwell
theory: the corresponding term is proportional to the NUT parameter and
it vanishes if $N=0$. Thus, in the starting solution this Dirac string
peculiarity is removable. From Eqs. (\ref{4.11}) and (\ref{4.18}) it
follows that the Dirac string for the metric (i.e. for the magnitude
$\vec V_1$) is absent if one imposes the restriction
\be\label{4.21}
\tilde\Q^T_IT_{II}=T^T_I\tilde\Q_{II},
\ee
on the starting charge configuration and the operator of the symmetry
transformation. Here $\tilde\Q_I$ and $\tilde\Q_{II}$ are respectively
$2\times 1$ and $2\times(2k+1)$ block components of the charge matrix
$\tilde\Q=\left(\tilde\Q_I\,\,\tilde\Q_{II}\right)$ which realizes a
real matrix representation of the complex charge parameter $\tilde q$.
Notice that all the relations which involve the matrix $\tilde\Q$ can
be obtained from the relations for $\tilde\Z$ by replacing
$\z_{\P}\longrightarrow q_{\P}$ in $\tilde\Z$; notice that we
have used a decomposition of $\tilde\Q$ similar to that of the Eqs.
(\ref{4.1})--(\ref{4.2})).

It is possible to solve the algebraic
restriction (\ref{4.21}) for the general case. However, for the
special case (\ref{4.12}), when the symmetry operator can be
represented in complex form, this can be done in aN easy and elegant
way. Actually, a simple algebraic analysis shows that in this case Eq.
(\ref{4.21}) leads to $N=0$ and the relations
\be\label{4.22}
e_p=r^{'}_pM, \quad\qquad g_p=r^{''}_pM,
\ee
i.e., to the NUT--less starting solution and to hard relations between
the electromagnetic charges and the non--electromagnetic sector of the
symmetry operator $T$. In this special case, up
to construction, the resulting metric is asymptotically flat at spatial
infinity. Notice that the resulting matter fields of heterotic string
theory also contain Dirac strings. In order to remove them one must
impose the corresponding restrictions on the magnitudes $\vec V_2$
and $\vec V_3$ (to elliminate, in turn, the terms proportional to
$\cos\theta$). Here we will not discuss these pure algebraic topics;
thus, our final solutions will include, for instance, magnetically
charged field configurations.

At the end of this Section let us note that our solution--generating
procedure, based on Eq. (\ref{3.12}), maps the full
Einstein--multi-Maxwell theory into the pure bosonic string theory
sector in the case $l_a\equiv 0$. Actually, this last restriction can
be imposed independently of our generation scheme. This fact, is also
reflected in the number of compactified space--time dimensions, which
is equal to $2k+1$, where $k$ is precisely the number of starting
Maxwell fields. Thus, a surprising fact is that the Abelian gauge field
sector of heterotic string theory is not related to the Maxwell sector
of the starting Einstein--multi-Maxwell theory: all the string theory
gauge fields depend only on the structure of the symmetry operator $T$.
Namely, one obtains $n$ Abelian vector fields $U(1)$ if one chooses the
parameter $l_a$ of height $n$.

%55555555555555555555555555555555555555555555555555555555555555555555555555555
\section{Conclusion and Discussion}
\setcounter{equation}{0}

The main result of this paper is the presentation of a new and explicit
scheme of generation of heterotic string theory solutions from stationary
fields of the Einstein--multi-Maxwell theory. Namely, one can start
with an arbitrary stationary solution of Einstein theory coupled to $k$
Maxwell fields and obtain a solution of heterotic string theory with
$n$ Abelian gauge fields which lives in $2(k+2)$ dimensions by making
use of the procedure developed in this paper. It is worth noticing that
our symmetry approach is based on pure algebraic calculations only as
far as all the potentials of the starting Einstein--multi-Maxwell
theory have been already computed.

Let us make two remarks concerning the properties of the underlying
symmetry map (\ref{3.12}) (or (\ref{4.5})--(\ref{4.7}) and (\ref{4.11})
in an equivalent and physically meaningful form). First of all, it is
interesting to notice that for the case $k=3$, the resulting heterotic
string theory becomes ten--dimensional. However, the complete theory
with $k=3$ (we refer to a theory with arbitrary potential $\tilde\Z$
of dimension $2\times8$, not to the Einstein theory with $3$ Maxwell
fields) corresponds to the bosonic sector of $N=4$ supergravity in
four dimensions. Thus, when $n=16$, our procedure relates $N=D=4$
and $N=1$, $D=10$ supergravities in a transparent form. Keeping this
in mind, it will be interesting to study the problem of supersymmetric,
and therefore BPS saturated, solutions in the framework of the
established correspondence. Namely, a question that  naturally arises
is: whether or not supersymmetric solutions of the four dimensional
theory map into supersymmetric solutions of the ten--dimensional one.
If they do so, how many supersymmetries will preserve under this
correspondence? Notice that some classes of four--dimensional
supersymmetric solutions have been extensively studied during last
several years (see, for instance, \cite{bko}, \cite{ap}); some
special classes of ten--dimensional supersymmetric solutions have also
been obtained (see \cite{bbc}, \cite{youm}) and the topic is still
under active investigation till now.

Our second remark is related to the level of generality of the map
(\ref{3.12}). We consider this issue in the framework of
asymptotically flat field configurations in the three--dimensional
sense. Namely, we consider that the fields which are encoded in the
potential $\Z$ vanish at spatial infinity. In this sense our map
(\ref{3.12}) is complete with respect to the total group of
three--dimensional charging symmetries, i.e., to the transformations
that preserve the asymptotical triviality of the
matrix potential $\Z$. Thus, if in Eq. (\ref{3.12})
the potential $\tilde\Z$ and the
symmetry operator $T$ have the most general form, our procedure is
non--generalizable by making use of hidden symmetries that act in the
subspace of three--dimensional asymptotically flat field solutions.
However, if one starts with the potential $\tilde\Z$ and the symmetry
operator $T$ given in the matrix representation which corresponds to
the complex $\tilde\z$ and $t$, one partially loses the charging
symmetry
self--invariance of the resulting solutions of the heterotic string
theory.
In fact, one loses the part of the total charging symmetry group
of the heterotic string theory which breaks the special (complex)
structure of the matrix potential $\tilde\Z$ and the symmetry
operator $T$ given by Eqs. (\ref{4.2}) and (\ref{4.12}).
This lost symmetry sector is evidently
non--trivial and can be used for the further generalization of the
solutions obtained in the framework of the pure complex generating
scheme developed at the end of the previous Section.

In this paper we have constructed as well the string theory extension
of the Kerr--multi-Newman solution of the Einstein--multi-Maxwell
theory.
This extension was presented as some simple and natural application of
the developed general formalism. It was also shown how to remove all
the Dirac string peculiarities from the resulting multi--dimensional
metric field, so that the resulting space--time of heterotic string
theory is asymptotically flat. Note that the constructed family of
solutions constitutes the first example of extension of the
Kerr--multi-Newman solution to the realm of the heterotic string
theory.
The obtained class of solutions is really interesting from the point
of view of black hole physics and can be studied in detail in a
conventional manner \cite{youm}, \cite{bh}--\cite{ap}.

%GGGGGGGGGGGGGGGGGGGGGGGGGGGGGGGGGGGGGGGGGGGGGGGGGGGGGGGGGGGGGGGGGGGGG

\section*{Acknowledgments}
The work of both authors was supported by CONACyT grant $No.$\,\,$J34245-E$;
the work of A.H-A. was also supported by grant CIC-$4.18$ whereas the work of
O.V.K. - by grant RFBR $No.$\,\,$00\,02\,17135$. O.V.K. thanks IFM--UMSNH
for facilities and hospitality provided during his stay at Morelia,
Michoacan, Mexico.

%LLLLLLLLLLLLLLLLLLLLLLLLLLLLLLLLLLLLLLLLLLLLLLLLLLLLLLLLLLLLLL


\begin{thebibliography}{143}

\bibitem{gsw}
M.B. Green, J.H. Schwarz, E. Witten, ``Superstring theory'',
Cambridge Univ. Pr. (1987).

\bibitem{kir}
E. Kiritsis,
``Introduction to superstring theory'',
Leuven Univ. Pr. (1998).

\bibitem{youm}
D. Youm,
Phys. Rept. {\bf 316} (1999) 1.

\bibitem{hs}
S.F. Hassan, A. Sen,
Nucl. Phys. {\bf B375} (1992) 103.

\bibitem{ms}
J. Maharana, J.H. Schwarz,
Nucl. Phys. {\bf B390} (1993) 3.

\bibitem{marcus}
N. Marcus, J.H. Schwarz,
Nucl. Phys. {\bf B228} (1983) 145.

\bibitem{sen3}
A. Sen,
Nucl. Phys. {\bf B434} (1995) 179.

\bibitem{bgm}
P. Breitenlohner, D. Maison, G.W. Gibbons,
Commun. Math. Phys. {\bf 120} (1988) 295.

\bibitem{mep}
A. Herrera-Aguilar, O.V. Kechkin,
Int. J. Mod. Phys. {\bf A13} (1998) 393.

\bibitem{cs}
A. Herrera-Aguilar, O.V. Kechkin,
Phys. Rev. {\bf D59} (1999) 124006.

\bibitem{zf}
O.V. Kechkin,
``3D heterotic string theory: new approach and extremal solutions'',
to be published in Phys. Rev. {\bf D}, hep-th/0110206.

\bibitem{ernst}
F. J. Ernst,
Phys. Rev. {\bf 167} 5 (1968) 1175.

\bibitem{maz}
P.O. Mazur,
Acta Phys. Pol. {\bf B14} (1983) 219.

\bibitem{exact}
D. Kramer, H. Stephani, M. MacCallum, E. Herlt,
``Exact solutions of the Einstein field equations'',
Deutcher Verlag der Wissenschaften, Berlin, (1980).

\bibitem{m1}
A. Herrera-Aguilar, O.V. Kechkin,``String Theory Extensions of
Einstein--Maxwell Fields: the static case", to appear in
Int.J.Mod.Phys. {\bf A}.

\bibitem{iwp}
W. Israel, G.A. Wilson, J. Math. Phys. {\bf 13} (1972) 865;
Z. Perj\'es, Phys. Rev. Lett. {\bf 27} (1971) 1668.

\bibitem{bh}
V.P. Frolov, I.D. Novikov, ``Black hole physics: basic concepts and
new developments", Kluwer Acad. Pr. (1998).

\bibitem{ap}
A.W. Peet,
``TASI Lectures on Black Holes in String Theory", in `Boudler
1999, Strings, branes and gravity', 353; hep--th/0008241.

\bibitem{bko}
E. Bergshoeff, R. Kallosh, T.Ortin,
Nucl. Phys. {\bf B478} (1996) 156.

\bibitem{bbc}
I. Bakas, M. Bourdeau, G. Lopes Cardoso,
Nucl. Phys. {\bf B510} (1998) 103.

\end{thebibliography}
\end{document}